\documentclass[twocolumn,showpacs,preprintnumbers]{revtex4}
\usepackage{graphics}
\usepackage{graphicx}
\usepackage{dcolumn}
\usepackage{bm}
\usepackage{amssymb}

\begin{document}

\title{Mixed pairing symmetry in $\kappa$-(BEDT-TTF)$_2$X organic
superconductors from ultrasonic velocity measurements}
\author{Maxime Dion}
\author{David Fournier}\thanks{Present address: Department of Physics and Astronomy,
University of British Columbia, Vancouver, BC, Canada V6T 1Z4}
\author{Mario Poirier}
\author{Kim D. Truong}
\author{A.-M. S. Tremblay}
\affiliation{Regroupement Qu\'eb\'ecois sur les Mat\'eriaux de Pointe, D\'epartement de
Physique, Universit\'e de Sherbrooke, Sherbrooke, Qu\'ebec,Canada J1K 2R1}
\date{\today}

\begin{abstract}
Discontinuities in elastic constants are detected at the
superconducting transition of layered organic conductors $\kappa $-(BEDT-TTF)%
$_{2}$X by longitudinal and transverse ultrasonic velocity
measurements. Symmetry arguments show that discontinuities in shear elastic
constants can be explained in the orthorhombic compound only if
the superconducting order parameter has a mixed character that
can be of two types, either $A_{1g}+B_{1g}$ or $B_{2g}+B_{3g}$ in
the classification of irreducible
representations of the orthorhombic point group $D_{2h}$.
Consistency with other measurements suggests that the
$A_{1g}+B_{1g}$ ($d_{xy}+d_{z(x+y)}$) possibility is realized.
Such clear symmetry-imposed signatures of mixed order parameters
have not been observed in other superconducting compounds.
\end{abstract}

\maketitle



Unconventional, non s-wave, superconductors in solids seem
ubiquitously associated with strong electronic correlations. This
is the case in a wide variety of compounds that include heavy
fermions, ruthenates, cuprates as well as quasi-two-dimensional
half-filled organic charge transfer salts $\kappa $-(ET)$_{2}$X
(ET = BEDT-TTF) \cite{POWELL}. In most cases gaps with nodes
are observed, but the exact symmetry of the unconventional
superconducting order parameter is uncontroversial only in the
cuprates.

In this letter, we focus on the layered organics that exhibit antiferromagnetism 
and Mott insulating behavior, as the cuprates, and establish
the two-component nature of the singlet order parameter in the
orthorhombic compound $\kappa $-(ET)$_{2}$Cu[N(CN)$_{2}$]Br.
Previous studies suggest $d$-wave pairing with nodes, although $s$-wave symmetry
is sometimes seen. Measurements
sensitive to the $\vec{k}$-space dispersion, such as scanning
tunneling spectroscopy \cite{Arai} and thermal conductivity
\cite{Izawa}, favor $d_{xy}$ symmetry, namely nodes along the
nearest neighbor bonds (or equivalenty, between the orthorhombic
axes). Moreover, theoretical calculations based either on
spin-fluctuation mediated superconductivity
\cite{Kuroki,Li,Schmalian,Kondo} or on quantum cluster methods \cite%
{Kyung,Sahebsara} and variational approaches \cite{Watanabe} for
the Hubbard model, support the anisotropic $d$-wave picture with a
prevailing $d_{x^{2}-y^{2}}$ symmetry. Nevertheless, none of these
calculations has considered interlayer hopping, which, as we will
show, is necessary to explain the experimental data that we
present.

The ultrasonic probe is extremely sensitive to gap anisotropies
as the attenuation and velocity depend on both the direction of
wave propagation and the direction of polarization. Attenuation
experiments on UPt$_{3}$ \cite{Ellman1996,Shivaram1986} and of
Sr$_{2}$RuO$_{4}$ \cite{Lupien2001} perfectly illustrate how the
unconventional gap structure can be unraveled by such a versatile
technique. In organic charge transfer salts however, attenuation
experiments are hampered by the small size and the shape of single
crystals. Nevertheless, one experiment was successful for the
$\kappa $-(ET)$_{2}$Cu[N(CN)$_{2}$]Br compound \cite{Fournier},
but the interpretation of the results was complicated by a phase
separation occurring even in highly ordered samples.
Notwithstanding these difficulties, ultrasound \textit{velocity}
can also be used to obtain insights into the nature of the 
superconducting (SC) state in layered organics. 
Lattice anomalies \cite{Muller} and
elastic constant changes \cite{Frikach1,Simizu1} have been
identified, but no consistent effort has been yet dedicated to
identify the SC order symmetry.

We report anomalies observed at the SC transition temperature
$T_{c}$ on three elastic constants of monoclinic $\kappa
$-(ET)$_{2}$Cu(NCS)$_{2}$ and of orthorhombic $\kappa
$-(ET)$_{2}$Cu[N(CN)]$_{2}$Br. Even though these compounds belong
to different point groups, we expect similarities in the SC order
parameters because of their nearly identical electronic properties. To
understand discontinuities in elastic constants one can invoke
Landau-Ginzburg arguments \cite{Testardi,Levy} or perform
detailed BSC type calculations \cite{Kataoka,Wakai}. Since we
focus on symmetry properties, a Ginzburg-Landau (GL) approach
will suffice \cite{Walker,Sigrist,Millis,Testardi1}.

We use an acoustic interferometer \cite{Fournier} to measure
relative changes in velocity $\Delta V/V$ that allow us to
extract the corresponding relative variations in the elastic
constants $C$ through $\Delta C/C=2\Delta V/V$. The $\kappa
$-(ET)$_{2}$X crystals grow as platelets containing the highly
conducting planes whose normal is oriented along $\vec{a}^{\ast}$
for monoclinic $\kappa $-(ET)$_{2}$Cu(NCS)$_{2}$ and along
$\vec{b}$ for the orthorhombic $\kappa
$-(ET)$_{2}$Cu[N(CN)$_{2}$]Br. Thus, ultrasonic plane waves can
be propagated only along these normal directions. Pure
longitudinal and transverse waves cannot be propagated along the $\vec{a}%
^{\ast }$ axis of the monoclinic structure so, strictly speaking,
it is not possible to measure the $C_{ij}$'s individually as it
is the case for the orthorhombic material \cite{Dieulesaint}. However,
given the layered structure and the $\vec{a}$ axis orientation of
about $110^{\circ}$ instead of $90^{\circ}$ from the plane, we
neglect, as a first approximation, the off-diagonal elements of
the $C_{ij}$ matrix that differentiate the monoclinic structure
from the orthorhombic one. This simplifies the data treatment
without affecting the conclusions. With this approximation the
measured $C_{ij}$'s are given in Table~\ref{tab.1} for each
crystal.

\begin{table}[tbp]
\centering
\begin{tabular}{c|c|c}
Waves & Cu(NCS)$_2$ & Cu[N(CN)$_2$]Br \\ \hline\hline
Longitudinal & $C_{11}$ ($\vec{a^*}$) & $C_{22}$ ($\vec{b}$) \\
Transverse & $C_{55}$ ($\vec{c}$) & $C_{66}$ ($\vec{a}$) \\
Transverse & $C_{66}$ ($\vec{b}$) & $C_{44}$ ($\vec{c}$)%
\end{tabular}%
\caption{Elastic constants $C_{ij}$ with the appropriate polarisation of the
ultrasonic waves for two $\protect\kappa$-(ET)$_2$X compounds.}
\label{tab.1}
\end{table}

\begin{figure}[h]
\includegraphics[width=7cm]{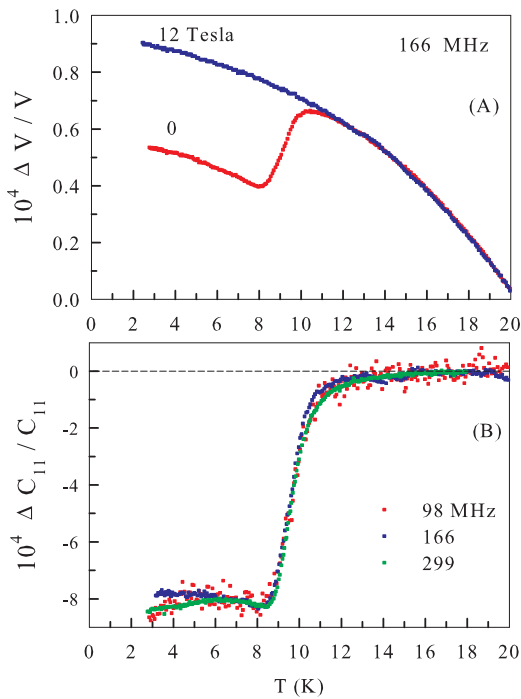}
\caption{(Color online) Longitudinal waves propagating along the
$\vec{a}^{\ast}$ axis in $\kappa$-(ET)$_{2}$Cu(NCS)$_{2}$: (A)
$\Delta V/V$ data at 166 MHz for H = 0 and 12 Tesla ; (B) $\Delta
C_{11}/C_{11}$ at three frequencies.} \label{fig.1}
\end{figure}

\begin{figure}[h]
\includegraphics[width=7cm]{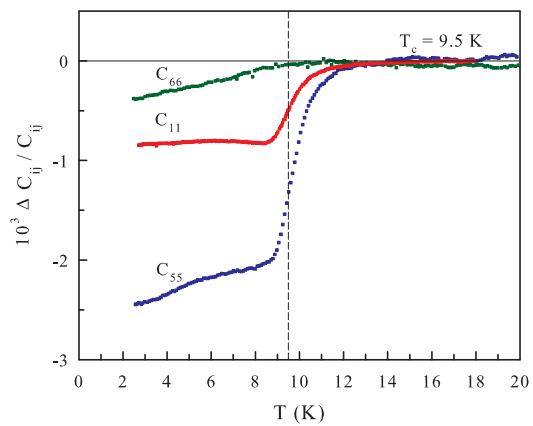}
\caption{(Color online) Temperature dependence of ${\Delta
C_{ij}}/C_{ij}$ for the $\kappa$-(ET)$_2$Cu(NCS)$_2$ compound.
The vertical dashed line indicates the SC critical temperature.}
\label{fig.2}
\end{figure}

The $\kappa $-(ET)$_{2}$Cu(NCS)$_{2}$ crystal will be considered
as our reference compound since it is located far enough from the
Mott transition line on the high pressure side of the P-T diagram
with no indication of a phase separation. To extract the elastic
change caused by the onset of superconductivity, we applied a
magnetic field perpendicular to the highly conducting plane to
quench the SC state. We show in Fig.~\ref{fig.1}A the temperature
dependence of the relative change of the longitudinal velocity
below 20 K at 166 MHz. In zero magnetic field a negative
discontinuity of the velocity is obtained at $T_{c}$ = 9.5 K; the
anomaly is completely quenched in a field of 12 Tesla leaving
only a monotonous decrease of the velocity as the temperature
increases. We notice the absence of magnetic field effects above
12 K, an observation that excludes, contrary to the $\kappa
$-(ET)$_{2} $Cu[N(CN)$_{2}$]Br compound \cite{Fournier}, the
presence of a coexisting phase in this temperature range. The
difference between these two curves yields the relative variation
of the compressional constant $C_{11}$ shown in Fig.~\ref{fig.1}B
at different frequencies. As expected, no frequency dependence is
observed: the onset of the SC phase yields a negative
discontinuity at $T_{c}$ that extends over a few degrees due to
important SC fluctuations above and below the superconducting
temperature defined as the maximum slope. At lower temperatures
${\Delta C_{11}}/C_{11}$ is practically constant. A similar
procedure was used for the two transverse acoustic modes
yielding, over the same temperature range, ${\Delta
C_{55}}/C_{55}$ and ${\Delta C_{66}}/C_{66}$. The three relative
elastic constant variations are compared in Fig.~\ref{fig.2}.
While a negative discontinuity is expected on $C_{11}$
\cite{Testardi}, the appearance of a discontinuity on the shear
constant $C_{55}$ is unusual. The amplitude of the discontinuity
is larger than that of $C_{11}$ by approximately a factor two,
excluding the simple explanation of mode mixing for a
quasi-transverse wave. These discontinuities are larger than in other non
conventional superconductors \cite{Lupien2,Bruls1990} by two to
three orders of magnitude. No discontinuity is observed for ${\Delta
C_{66}}/C_{66}$; only a small change of slope is obtained at
$T_{c}$. 

\begin{figure}[h]
\includegraphics[width=7cm]{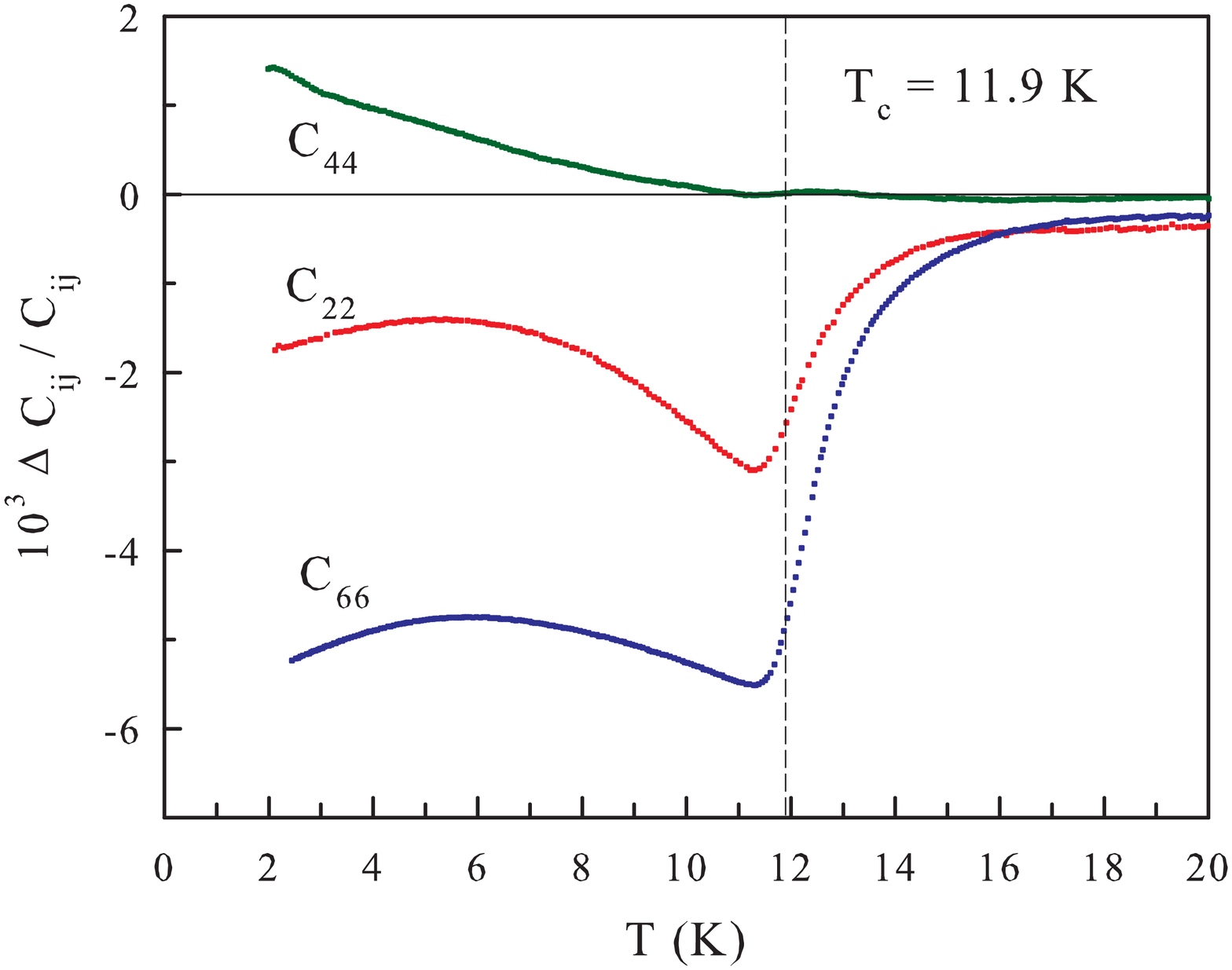}
\caption{(Color online) Temperature dependence of ${\Delta
C_{ij}}/C_{ij}$ for the $\kappa $-(ET)$_{2}$Cu[N(CN)$_{2}$]Br
compound obtained with a 16 Tesla magnetic field. The dashed line
indicates the SC critical temperature.} \label{fig.3}
\end{figure}

In the highly ordered $\kappa $-(ET)$_{2}$Cu[N(CN)$_{2}$]Br
compound, the coexistence of antiferromagnetic (AF) and SC phases
complicates the analysis of the SC state \cite{Fournier}.
Moreover, a higher magnetic field is needed to quench the SC
state. We present in Fig.~\ref{fig.3} the $\Delta C_{ij}/C_{ij} $
obtained by substracting the zero and 16 Tesla curves. We notice
that magnetic field effects are observed in the normal state up
to 20 K on $C_{22}$ and $C_{66}$ ($\Delta C_{ij}/C_{ij}$ is not
zero). The temperature dependence below $T_{c}$ = 11.9 K is also
not monotonous and the SC fluctuations appear on a wider
temperature range above $T_{c}$. Notwithstanding these
differences, the comparison with the $\kappa
$-(ET)$_{2}$Cu(NCS)$_{2}$ data (see Fig.~\ref{fig.2}) at $T_{c}$
is remarkable: we still observe a negative discontinuity on
$C_{22}$ ($C_{11}$), a larger one on $C_{66}$ ($C_{55}$) and only
a change of slope on $C_{44}$ ($C_{66}$). These observations
clearly establish the similarity of the couplings between the SC
order parameter and the elastic strains, although the crystal
symmetry groups differ because of the tilting of the axis normal
to the planes. Moreover, they confirm that the negative
dicontinuity on ${\Delta C_{55}}/C_{55}$ for the monoclinic
compound is intrinsic and that it cannot be attributed to mode
mixing.

Experiment has established that the layered organics are singlet
superconductors \cite{POWELL}. In the simplest GL model then,
discontinuities in elastic constants at the superconducting
transition are easily explained through the free energy functional
\begin{equation}
F=a|\eta |^{2}+g\varepsilon _{i}|\eta |^{2}+\frac{b}{2}|\eta
|^{4}+ \sum_{i,j} \frac{1}{2}C_{ij}\,\varepsilon _{i}\varepsilon
_{j} \label{GL_simple}
\end{equation}
where $\eta$ is the order parameter, $b$ is a constant,
$\varepsilon _{i}$ is the strain, $C_{ij}$ the matrix of elastic
constants, while $a$ is proportional to ($T-T_{c}$). If one of
the strains is coupled linearly through the constant $g$ to the
order parameter, the minimization with respect to $\eta $ shows
that at the transition a negative discontinuity appears on the
effective elastic constant $C_{ii}^{\prime }=\partial
^{2}F/\partial \varepsilon _{i}^{2}$. Such a linear coupling to
$|\eta |^{2}$ is possible only if the strain $\varepsilon _{i}$
is invariant under all the operations of the point group because
$|\eta |^{2}$ is. Higher order coupling terms in the
free energy would only lead to the change of slope or curvature
observed below $T_{c}$ for all $C_{ii}$, and these are not
considered here.

Table \ref{tab:groupe_C2_Cu} shows a simplified character table for the
irreducible representations of the monoclinic $C_{2h}$ group of $\kappa $%
-(ET)$_{2}$Cu(NCS)$_{2}$, along with the transformation
properties of the strains and examples of basis functions for the
order parameter. Note that the $x$ and $y$ axis are not
perpendicular. They lie along the atomic bonds, which are along
the diagonal formed by the $b$ and $c$ axes. Since, according to
Table \ref{tab:groupe_C2_Cu}, $\varepsilon _{1}$ and $\varepsilon
_{5}$ are invariant under the symmetry operations of the
group, the corresponding elastic constants can couple linearly to
$|\eta |^{2}$, leading to negative discontinuities. However,
$\varepsilon _{6}$ is not invariant so there is no discontinuity
at $T_{c}.$ This explains the observations for $\kappa
$-(ET)$_{2}$Cu(NCS)$_{2}$ and it does not impose any constraint on
the symmetry of the order parameter.
\begin{table}[tbp]
\centering
\begin{tabular}{c||cc||c|c}
irrep & $E$ & $C_2^\mathbf{b}$ & Basis functions & Strains \\
\hline\hline
$A_g$ & 1 & 1 & $s$, $xy$, $(x+y)z$ & $\epsilon_1$, $\epsilon_2$, $%
\epsilon_3 $, $\epsilon_5$ \\
$B_g$ & 1 & -1 & $x^2-y^2$, $(x-y)z$ & $\epsilon_4$, $\epsilon_6$%
\end{tabular}%
\caption{Simplified character table, basis functions and transformation
properties of the strains. The monoclinic $C_{2h}$ group for $%
\protect\kappa $-(ET)$_{2}$Cu(NCS)$_{2}$ has the character table of $C_2 \otimes i$,
but inversion $i$ always has character $+1$ for singlets so the table of $C_2$
shown above suffices. The names of the irreducible representations
are those of $C_{2h}$. The last column shows the transformation properties of the
strains and the next to last column examples of basis functions for the
order parameter. The $\mathbf{a}$ axis is tilted towards $\mathbf{c}$
(equivalently $x+y$) axis in the layers.}
\label{tab:groupe_C2_Cu}
\end{table}

In the orthorhombic $\kappa $-(ET)$_{2}$Cu[N(CN)$_{2}$]Br,
because of the different conventions, the role of $\epsilon _{5}$
in the monoclinic case is played by $\epsilon _{6}$. The
simplified character table \ref{tab:groupe_D2h_Br} for the
$D_{2h}$ group shows that the shear strain $\epsilon _{6}$ is not
invariant under the operations of the group.
\begin{table}[tbp]
\centering
\begin{tabular}{c||cccc||c|c}
irrep & $E$ & $C_2^\mathbf{a}$ & $C_2^\mathbf{b}$ & $C_2^\mathbf{c}$ &
Basis fcts & Strains \\ \hline\hline
$A_{1g}$ & 1 & 1 & 1 & 1 & $s$, $xy$ & $\epsilon_1$, $\epsilon_2$, $%
\epsilon_3 $ \\
$B_{1g}$ & 1 & -1 & -1 & 1 & $(x+y)z$ & $\epsilon_6$ \\
$B_{2g}$ & 1 & -1 & 1 & -1 & $x^2-y^2$ & $\epsilon_5$ \\
$B_{3g}$ & 1 & 1 & -1 & -1 & $(x-y)z$ & $\epsilon_4$%
\end{tabular}%
\caption{Simplified character table, basis functions and transformation
properties of the strains for the orthorhombic $D_{2h}=D_2\otimes i$ group
appropriate for $\protect\kappa $-(ET)$_{2}$Cu[N(CN)$_{2}$]Br. The $\mathbf{b%
}$ axis is perpendicular to the layers and the $x+y$ axis is along
$\mathbf{a}$.}
\label{tab:groupe_D2h_Br}
\end{table}
Hence, the $C_{66}$ negative discontinuity at $T_{c}$ cannot be
explained with the simplest model Eq.(\ref{GL_simple}). One must
introduce an order parameter with two orthonormal basis
functions with respective complex coefficients $\eta _{1}$ and
$\eta _{2}$. Let us first neglect the strain terms and consider
the most general free energy functional that is invariant under
the point group and phase changes of the order parameter
\cite{Sahu}
\begin{eqnarray}
F_{\eta } &=&a_{1}|\eta _{1}|^{2}+\frac{b_{1}}{2}|\eta
_{1}|^{4}+a_{2}|\eta _{2}|^{2}+\frac{b_{2}}{2}|\eta _{2}|^{4}
\label{GL_general} \nonumber \\  &&+|\eta _{1}|^{2}|\eta
_{2}|^{2}(\gamma +\delta \cos (2\Delta \theta )).
\end{eqnarray}%
In this expression, $\gamma $ and $\delta $ are constants and $\Delta \theta
$ is the phase difference between the two components of the order parameter.
If $\delta $ is positive, this free energy will be minimized by $\Delta
\theta =\pm \pi /2,$ while if $\delta $ is negative $\Delta \theta =0$ or $%
\pi $ will be the minimum. The case $\Delta \theta =\pm \pi /2$ corresponds
to a complex order parameter, hence it breaks time reversal symmetry.

To explain the discontinuity in the transverse elastic constant,
the coupling free energy
\begin{equation}
F_{\eta \varepsilon }=g\epsilon _{6}|\eta _{1}||\eta _{2}|\cos (\Delta
\theta )
\end{equation}
must be allowed by symmetry. Also, $\cos (\Delta \theta )$ should
not vanish, thus removing the possibility of a time-reversal
symmetry-breaking state. Since $\varepsilon _{6}$ transforms
according to the $B_{1g}$ representation, there are only two
possibilities. Either one of the $\eta $ is invariant ($A_{1g}$)
and the other one transforms as $B_{1g}$ or one of the components
transforms like $B_{2g}$ and the other one like $B_{3g}$. This can
be checked by showing that the product of the characters in Table
\ref{tab:groupe_D2h_Br} is unity for all group operations applied
to $F_{\eta \varepsilon }$. Note that both of the above
possibilities for $\eta _{1}$ and $\eta _{2}$ forbid a linear
coupling to $\varepsilon _{4}$ since the latter transforms like
$B_{3g}$. This explains the absence of a discontinuity in the
corresponding elastic constant.

Since both scanning tunneling spectroscopy \cite{Arai} and thermal
conductivity \cite{Izawa} suggest nodes along the $x$ and $y$
axis, this forces us to choose an order parameter that has a
mixed $A_{1g}+B_{1g}$ character, namely $d_{xy}+d_{z(x+y)}$. The
nodeless $s$ case has the same symmetry as $d_{xy}$ so more
generally it should be included but it suffices that its
amplitude be smaller than that of $d_{xy}$ for the nodes of
$s+d_{xy}$ to survive. They are just shifted from their position
in the $d_{xy}$ case. The $d_{z(x+y)}$ component does not remove
the nodes in the plane, but it clearly breaks mirror symmetry
about the planes.

On general grounds, the free energy Eq.(\ref{GL_general})
predicts two different $T_{c}$'s since there is no a priori reason
why $a_{1}$ and $a_{2}$ should vanish at the same $T$. That is
different from the case of Sr$_{2}$RuO$_{4}$ where the two
components of the order parameter necessary to explain the data
belong to a single two-dimensional representation $E_{2u}$ of the
point group $D_{4h}$ \cite{Walker}. Although the present lattice
is nearly triangular, the two components of the order parameter
that we found do not coalesce into a single two-dimensional
representation of the $D_{6h}$ group \cite{Kuznetsova}.
Nevertheless, the mixed $A_{1g}+B_{1g}$ representation for the
orthorhombic crystal does coalesce into the one-dimensional
$A_{g}$ representation of its monoclinic cousin, leading to a
single $T_{c}$ in that case. Hence, we do not expect a large
difference between the two transition temperatures of the
orthorhombic crystal. Our experimental data in Fig.~\ref{fig.3}
show a rather broad transition with an extended region of SC
fluctuations that could mask this difference between the two
transitions.

The presence of a $d_{z\left( x+y\right) }$ component to the order parameter
suggests that interlayer hopping is an important variable in the problem.
The value of this parameter has been estimated from angle-dependent
magnetoresistance oscillations \cite{Goddard}. Thermal expansion data \cite%
{deSouza} also disclose a striking anisotropy and dependence of $T_{c}$ on
interlayer effects \cite{Muller2} that are unlikely to be captured by a 2D
purely electronic model.

In summary, symmetry and the observed discontinuities at $T_{c}$
in the ultrasonic velocity data for two compounds of the layered
$\kappa $-(ET)$_{2}$X organic superconductors demonstrate that
the order parameter must have at least two components in the
orthorhombic compound $\kappa $-(ET)$_{2}$Cu[N(CN)$_{2}$]Br.
Consistency with other experiments selects $A_{1g}+B_{1g}$
(equivalently $d_{xy}+d_{z\left( x+y\right)}$). The two
components coalesce into a one-dimensional irreducible
representation $A_{g}$ in the monoclinic compound $\kappa
$-(ET)$_{2}$Cu(NCS)$_{2}$. Nodes are not symmetry imposed but are
symmetry allowed and are likely to occur in electronic pairing
mechanisms. The $d_{z\left( x+y\right) }$ component of the order
parameter suggests that further studies of interlayer coupling
are called for.

The authors achnowledge stimulating discussions with Claude
Bourbonnais, David S\'{e}n\'{e}chal and Peter Hirschfeld and they
thank Mario Castonguay for technical support. This work was
supported by grants from the Fonds Qu\'{e}b\'{e}cois de la
Recherche sur la Nature et les Technologies (FQRNT), from the
Natural Science and Engineering Research Council of Canada
(NSERC). A.-M.S.T. also acknowledges the support of the Tier I
Canada Research Chair program and of the Canadian Institute for
Advanced Research.

\end{document}